\documentstyle[epsfig,longtable]{aipproc}

\begin{document}
\title{The IBIS view of the galactic centre: INTEGRAL's imager observations
simulations}

\author{P. Goldoni$^*$, A. Goldwurm$^*$, P. Laurent$^*$, F. Lebrun$^*$}
\address{$^*$ CEA/DSM/DAPNIA, SAp CEA-Saclay F-91191 Gif-sur-Yvette FRANCE}

\maketitle

\begin{abstract}
The Imager on Board Integral Satellite (IBIS) is the imaging
instrument of the INTEGRAL satellite, the hard-X/soft-gamma
ray ESA mission to be launched in 2001. It provides diagnostic capabilities
of fine imaging (12' FWHM), source identification and spectral
sensitivity to both continuum and broad lines over a broad
(15~keV--10~MeV) energy range. It has a continuum sensitivity of 
2~10$^{-7}$~ph~cm$^{-2}$~s$^{-1}$ at 1~MeV for a 10$^6$ seconds observation and
a spectral resolution better than 7~$\%$ at 100~keV and of 6~$\%$ at 1~MeV.
The imaging capabilities of the IBIS are characterized by the
coupling of the above quoted source discrimination capability
with a very wide field of view (FOV), namely 9$^\circ$ $ \times $ 9$^\circ$ 
fully coded, 29$^\circ$ $ \times $ 29$^\circ$ partially coded FOV.

\noindent We present simulations of IBIS observations of the Galactic
Center based on the results of the SIGMA Galactic Center survey. 
They show the capabilities of this instrument in
discriminating between different sources while at the same time
monitoring a huge FOV. It will be possible to simultaneously take
spectra of all of these sources over the FOV even if the sensitivity
decreases out of the fully coded area. It is envisaged that a proper
exploitation of both the FOV dimension and the source localization
capability of the IBIS will be a key factor in maximizing its
scientific output.
\end{abstract}

\section*{The IBIS telescope}

The IBIS telescope \cite{pg2:Ubertini96} is a coded mask imaging system 
based on a 1.5 cm thick tungsten mask of 95 $\times$ 95 elements of 
1.1 $\times$ 1.1 cm$^2$, designed from a replication of a 53~$\times$~53 MURA 
basic pattern \cite{pg2:Gottesman89} which gives it a high angular
resolution ($\approx$ 12') over a wide field of view (FOV) (9$^\circ$
$ \times $ 9$^\circ$ fully coded, 29$^\circ$ $ \times $ 29$^\circ$
partially coded at zero response).
The IBIS detection system is composed of two planes, 
an upper layer made of 16384 squared CdTe pixels (ISGRI) and 
a lower layer made of 4096 CsI scintillation bars (PiCsIT). 
This system enables high sensitivity continuum spectroscopy (E/$\Delta $E$>$10)
and a wide spectral range (15 keV -- 10 MeV).

\noindent The simulation we performed are for the moment limited to the ISGRI 
upper layer\cite{pg2:Lebrun95}. The ISGRI pixels are 4$ \times $
4 mm$^2$, 2~mm thick crystals of Cadmium Telluride, a semiconductor
operating at ambient temperature, providing a spectral resolution of
about 8$\% $ at room temperature.
The 128$ \times $128 = 16384 pixels are arranged in 8 modules separated by 
dead zones 2 pixel wide needed by the mechanical structures which sustain
the detector plane. The sensitivity loss caused by dead zones is not large,
however the absence of sensitive elements in the detector plane must be
properly taken into account during deconvolution procedures. 


\begin{figure}[b!] 
\centerline{\epsfig{file=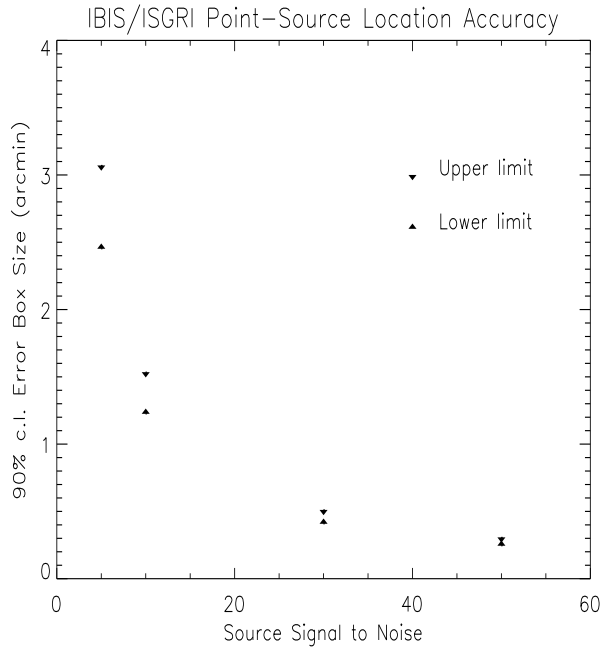,height=3.5in,width=3.5in}}
\vspace{10pt}
\caption{\it IBIS/ISGRI Point Source Location Accuracy vs. signal to noise. Values are comprised between triangles}
\label{F:pg2:1}
\end{figure}

\begin{figure}[b!] 
\centerline{\epsfig{file=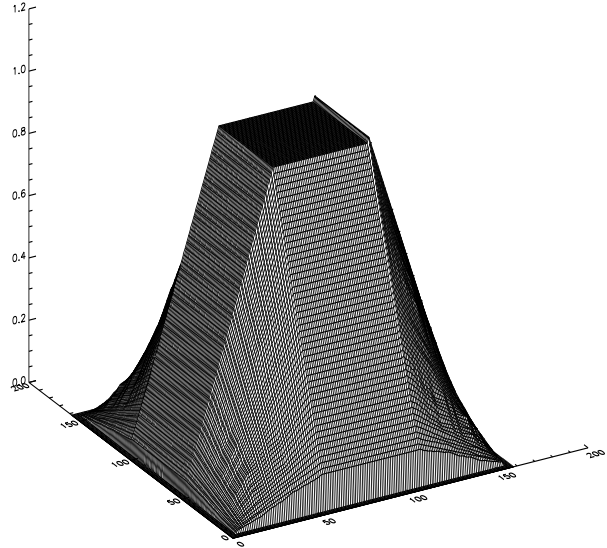,height=3.5in,width=3.5in}}
\vspace{10pt}
\caption{\it 3D image of the IBIS/ISGRI normalized sensitivity over the
whole field of view.}
\label{F:pg2:2}
\end{figure}

\section*{Coded Mask Imaging}
 
In the X/gamma-ray domain, source localization can be achieved through the use
of coded mask imaging systems \cite{pg2:Caroli87}. These systems represent
the most performant way to achieve source localization in this energy domain.
In recent years this was demonstrated by the fast localization of X-ray
Novae by the SIGMA coded mask telescope \cite{pg2:Vargas97}and by the first
localization of the X counterpart of a gamma ray burst thanks to the
Wide Field Cameras of the BeppoSAX satellite \cite{pg2:Costa97} .

\noindent The incoming radiation is modulated by a mask with opaque and transparent 
elements and then recorded by a position sensitive detector. 
In this way it is possible to simultaneously measure source and background 
thus avoiding the on/off technique usually employed in gamma-ray astronomy
which is very sensitive to background variability and observing conditions.
The angular resolution of such system is defined by the angle subtended by 
one hole at the detector which is 12' for IBIS. However point source
location accuracy will also depend on the ratio R between mask element
size and detector pixel size and will be proportional to the source
signal to noise (S/N) ratio. The positional error can be easily computed
for the case of integer ratios R as a function of S/N \cite{pg2:Goldwurm95}. 
ISGRI/IBIS pixels subtend an angle of $\sim$ 4.6' and therefore R is equal
to 2.4, we thus expect to reach a position accuracy which is between the
values obtained for R=2 and for R=3. These values for a number of S/Ns are
reported in Fig. 1. The instrument's extended field of view 
is divided in two parts, the fully coded field of view (FCFOV), where each
source will project a complete (shifted) basic pattern onto the detector plane,
and the partially coded FOV (PCFOV) where source flux will be only partially
modulated by the mask.
Telescope's sensitivity will depend on the amount of modulation 
and therefore it will be constant in the FCFOV and decreasing with angle from
telescope axis in the PCFOV. This is shown in Fig. 2. 

\noindent Sky images are produced through an algorithm which is basically a
balanced correlation between detector image and mask pattern (see \cite{pg2:Goldwurm95}). Reconstructed sky images however are affected by large amount of
coding noise in form of ghost peaks and extended modulation due to the sources 
in the FC or PC FOVs. An iterative source cleaning procedure must therefore been applied in the data analysis to correctly deconvolve sky images.
Such procedures have been developed by our group and succesfully used for 
the data analysis of the SIGMA coded mask telescope images
\cite{pg2:Goldwurm95} and now modified and adapted to simulated IBIS images.
Indeed the simulations
will allow us to test and improve the data analysis techniques.

\section*{Imaging simulations of the Galactic Bulge}

The Galactic Bulge is the zone of the sky with the highest source density in 
the X/gamma-ray domain.
A typical INTEGRAL Galactic Bulge observation will last about 10$^{5}$~s, 
i.e. a day. We simulated such an observation in the energy range 40-80~keV 
assuming sources are at their average brightness level as measured with
the SIGMA/GRANAT telescope \cite{pg2:Paul91} during its 7 year long 
galactic center survey (\cite{pg2:Goldwurm94,pg2:Vargas97}) and 
adding a contribution also from 4U~1700--37, a hard X-ray high mass
binary, at a level about 1/2 lower than the peak flux value detected by
SIGMA \cite{pg2:Laurent92}. We then applied sky image reconstruction and cleaning
algorithm as described in the previous section.

\noindent Results of the simulation (Fig. 3) show the expected imaging
performances of the IBIS telescope. All sources including the faint Terzan~1 
($\approx$~8~mCrabs) are clearly detected at more then 5 sigma level
with the brighter (the ``microquasar'' 1E~1740.7-2942) at more than 40 sigma. 
The sources are well separated even in the very center of the galaxy where
source confusion problems may arise.
It can be seen that 4U~1700--37 is also easily detected by the IBIS instrument,
which proves the possibility of taking a high significance spectrum even if 
the source is at more than 11$^{\circ}$ from pointing axis. The possibility to
monitor such huge field gives to IBIS a very high potential for search and 
high precision ($<$~1') localization of high-energy brigth transient sources
like X-ray novae and gamma-ray bursts.
Moreover thanks to the wide field and the galatic plane survey program, 
it will be possible to detect serendipitous,
faint sources like Ti$^{44}$ lines from hidden supernovae in the galactic
plane.


\begin{figure}[b!] 
\centerline{\epsfig{file=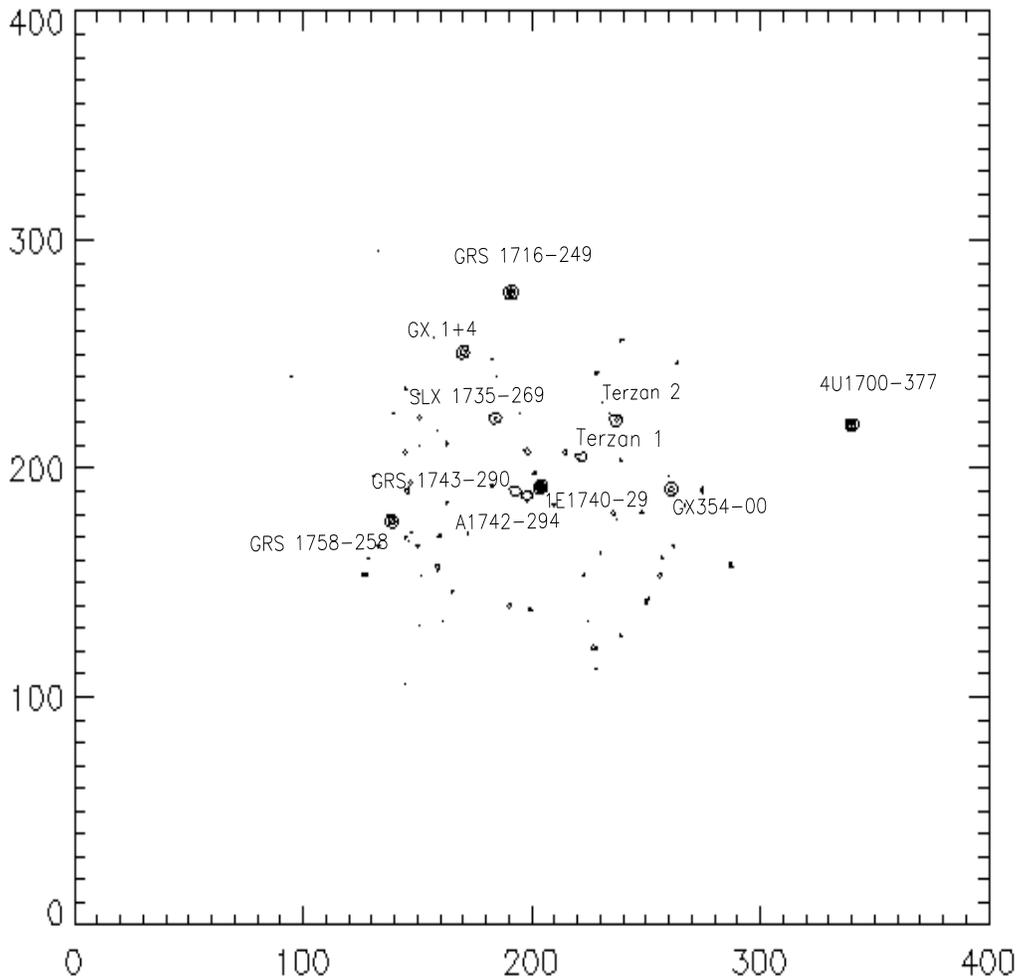,height=5.8in,width=6in}}
\vspace{10pt}
\caption{\it 5-level contour image of the complete ISGRI FOV of
the Galactic Center in the 40-80 keV band. Units on axis are sky
pixels (about 4.6 arcminutes see text). Note the clear detection of the 4U1700-377 flare at more than 11 $^{\circ}$ from the pointing direction}
\label{F:pg2:3}
\end{figure}

\end{document}